# An efficient slope stability algorithm with physically consistent parametrisation of slip surfaces

Leonardo Maria Lalicata[1], Andrea Bressan[2], Simone Pittaluga[3], Lorenzo Tamellini[2], Domenico Gallipoli[1]

1 Department of Civil, Chemical and Environmental Engineering, University of Genova

2 Consiglio Nazionale delle Ricerche, Istituto di Matematica Applicata e Tecnologie Informatiche "E. Magenes" (CNR-IMATI). Via Ferrata, 5/A, 27100, Pavia, Italy

3 Consiglio Nazionale delle Ricerche, Istituto di Matematica Applicata e Tecnologie Informatiche "E. Magenes" (CNR-IMATI). Via dei Marini 6, 16149, Genova, Italy

*Abstract*

This paper presents an optimised algorithm implementing the method of slices for analysing the stability of slopes. The algorithm adopts an improved physically based parameterisation of slip lines according to their geometrical characteristics at the endpoints, which facilitates the identification of all viable failure mechanisms while excluding unrealistic ones. The minimisation routine combines a preliminary discrete calculation of the factor of safety over a coarse grid covering the above parameter space with a subsequent continuous exploration of the most promising region via the simplex optimisation. This reduces computational time up to about 92% compared to conventional approaches that rely on the discrete calculation of the factor of safety over a fine grid covering the entire search space. Significant savings of computational time are observed with respect to recently published heuristic algorithms, which enable a continuous exploration of the entire parametric space. These efficiency gains are particularly advantageous for numerically demanding applications like, for example, the statistical assessment of slopes with uncertain mechanical, hydraulic and geometrical properties. The novel physically based parametrisation of the slip geometry and the adoption of a continuous local search allow exploration of parameter combinations that are necessarily neglected by standard grid-based approaches, leading to an average improvement in accuracy of about 5%.

*Keywords*: slope stability, Bishop method, simplex optimisation, efficient algorithm

# 1 Introduction

Failure of slopes can be catastrophic in terms of casualties and economic losses. In Italy, for example, the occurrence of landslides poses a hazard to many areas (Trigila *et al.*, 2010, Jaedicke *et al.*, 2013) as recently demonstrated by the slope instabilities in the Emilia Romagna region (ISPRA, 2023), the mudslides in the Ischia island (Romeo *et al.*, 2023) and the shallow and deep-seated slips in the Italian Alps and Apennines (Notti *et al.*, 2021; Cotecchia *et al.*, 2019; Pedone *et al.*, 2018). Landslide susceptibility maps constitute an effective tool for managing such hazards and are produced via either statistical (Govi *et al.*, 1985; Versace *et al.*, 2002) or deterministic methods (Cascini *et al.*, 2005; Montrasio and Valentino, 2008; Federici *et al.*, 2015; Rahardjo *et al.*, 2023). Deterministic methods have greater generality than statistical ones as they can account for local factors such as the hydro-mechanical and geometric characteristics of the slope. Nevertheless, to reduce the computational costs associated to the performance of a large number of calculations, these methods tend to employ simple infinite slope models, which are adequate for shallow slides but not for deeper rotational mechanisms (Duncan, 1996).

For rotational slides, the method of slices (Fellenius, 1936; Janbu, 1954, 1973; Bishop, 1955; Bishop and Morgenstern, 1960; Morgenstern and Price, 1965; Spencer, 1967, 1973; Bell, 1968; Sarma, 1973, 1979) has long been employed by engineers worldwide to calculate the safety factor and is implemented in many commercial and non-commercial software (e.g. SLOPE/W, 2007; SLIDE2; Steward *et al.*, 2011). These implementations adopt different algorithms to search for the critical slip surface. i.e. the slip surface with the lowest factor of safety, across the space of all potential failure mechanisms (Boutrop and Lovell, 1980; Siegel *et al.*, 1981; Baker, 1980; Nguyen, 1985; Celestino and Duncan, 1985, Zolfaghari *et al.*, 2005; Cheng *et al.*, 2007; Himanshu and Burman, 2019; Himansu *et al.*, 2021; Kalatehjari *et al.*, 2012). Among them, heuristic algorithms, such as genetic and particle swarm optimisation, have proven relatively accurate because they provide a continuous, rather than discrete, exploration of the search space while being less prone to converge towards local minima of the factor of safety. The main drawback of heuristic algorithms consists, however, in their high computational cost as they require a large number of calculations of the factor of safety for a single slope stability analysis. This is particularly problematic for the production of landslide susceptibility maps as the consideration of material and physical uncertainties at the regional scale often requires the use of computationally expensive statistical tools and, hence, the run of a large number of analyses (Le *et al.*, 2015).

In this context, the present work develops a highly efficient algorithm that overcomes the above limitations by incorporating: a) a novel unbiased parametrisation of slip surfaces identifying all and only viable rotational mechanisms and b) the combination of a discrete probe of the entire search domain with the continuous exploration of only the most promising region. Regarding the former point, slip lines are defined by geometric parameters at their endpoints, which are bound by physically based limits. This allows consideration of all viable configurations while excluding unrealistic ones to avoid unnecessary computational efforts. Regarding the second point, the proposed hybrid search drastically reduces computational time while also producing an increase of accuracy with respect to discrete (i.e. grid-based) routines.

The algorithm is illustrated here using the Bishop (1955) method of slices for circular slip surfaces, which is one of the limit equilibrium methods endorsed by Eurocode 7 (EN 1997-1) for the stability analysis of slopes (Bond *et al.*, 2013). For homogeneous slopes, the Bishop (1955) method produces results nearly identical to those obtained from more complex methods of slices, such as Morgenstern and Price (1965) and Spencer (1967), and deviates by less than 1.4%, on average, from the upper bound limit analyses of Huang (2023). This high accuracy, together with the compliance with existing norms and the simplicity of calculations, explains the popularity of Bishop (1955) method among engineering practitioners, justifying its use in the present study. Yet, the proposed algorithm has general applicability and can be adapted to other methods of slices with non-circular slip surfaces, regardless of the specific material and geometric characteristics of the slope.

The work presented in this paper constitutes the initial part of a wider research programme about the study of slope instabilities to produce landslide susceptibility maps accounting for geometrical and material uncertainties at regional scale.

## 2 Method of slices

With reference to a generic slope having height $H$, inclination $\beta$ and length $B = H/\tan\beta$ (Figure 1a), the method of slices calculates the factor of safety $F$ of a given slip surface as the ratio between the available resistance $R$ and the driving action $A$ acting on the potentially unstable soil mass:

$$F = \frac{R}{A} \tag{1}$$

The driving action $A$ is generated by the weight of the unstable mass while the available resistance $R$ is the maximum resistance allowed by the soil strength.

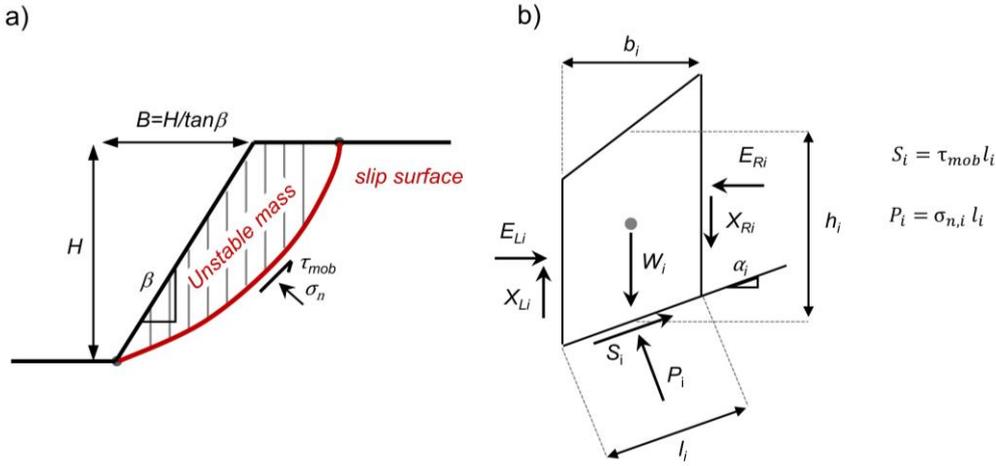

Figure 1. a) Schematic representation of the method of slices: a) potentially unstable soil mass cut into multiple vertical slices, b) distribution of forces for the $i^{th}$ slice.

To calculate the available resistance $R$, the problem is made statically determined by dividing the unstable mass in $n$ vertical slices (Figure 1a). The equilibrium of each slice is then solved after the enforcement of the soil failure criterion on the slip surface, which is loaded by the normal stress $\sigma_n$ and the mobilised shear strength $\tau_{mob}$, and the introduction of simplifying hypotheses about interslice forces.

This work focuses on algorithmic aspects and therefore assumes a simple homogeneous dry slope with cohesion $c$, friction angle $\varphi$, and unit weight $\gamma$. Among the different methods of slices, the Bishop (1955) method has been chosen because of its popularity among engineering practitioners, which is primarily due to the simplicity of calculations, good accuracy and compliance with design norms such as Eurocode 7 (EN 1997-1). Under the above hypotheses, the factor of safety $F$ of a given slip surface is calculated from Equation (1) according to Bishop (1955) as:

$$F = \frac{1}{\sum_{i=1}^{n} W_i \sin\alpha_i} \sum_{i=1}^{n} \frac{cb_i + W_i \tan\varphi}{\left(1 + \frac{\tan\alpha_i \tan\varphi}{F}\right)\cos\alpha_i} \tag{2}$$

where $b_i$ and $\alpha_i$ are respectively the width and the base angle of the $i^{th}$ slice of height $h_i$ while $W_i = \gamma h_i b_i$ is the corresponding weight (Figure 1b). The factor of safety $F$ appears both on the left- and right-hand sides of Equation (2), so it must be calculated via an iterative procedure. A choice of $n = 25$ equally spaced slices has been made as a good compromise between accuracy and computational costs.

The factor of safety of the slope coincides with the smallest value calculated by Equation (2) in correspondence of the critical slip surface. Therefore, from the mathematical viewpoint, the calculation of this factor of safety poses an optimisation problem consisting in the search for the global minimum of Equation (2) across the space of all possible circular failure mechanisms. This search typically requires the analysis of many slip geometries and necessitates a numerically efficient algorithm, especially if statistical models are used to describe material, geometrical and physical uncertainties.

## 3  Geometrical parametrisation of slip surfaces

Most implementations of Bishop (1955) method search for the critical slip surface across a large set of distinct circumference arcs, which is generated by changing the centre, $C$ over a fixed grid while varying, for each centre, the radius, $R$ over a fixed interval (Figure 2). In other implementations, the above set of circumference arcs is instead generated by changing the pair of entry and exit points into/from the soil while varying, for each pair, the radius over a fixed interval. The definition of the domain of centres and the interval of radii is, however, very much dependent on the user experience. In some cases, this might result in the definition of a set of slip lines excluding the critical mechanism, thus arriving to an incorrect result, and/or including mechanically unviable mechanisms, thus leading to unnecessary calculations. In addition to this, the selection of equally spaced radii does not allow a uniform sampling of the search domain as it will be shown later.

To overcome the above limitations, this paper proposes a physically based parametrisation of the slip surface, which enables to identify the set of circumference arcs that strictly encompasses all realistic geometries, with consequent gains of efficiency and accuracy. This novel parametrisation defines each slip surface via the abscissa of the entry point, $x_{in}$ and the abscissa of the exit point, $x_{out}$, as in previous works, but considers, as third parameter, the tangent angle of the slip line at the entry point, $\delta$ instead of the radius $R$ (Figure 2). This choice has the advantage of ensuring that all three parameters are bound by physically consistent limits, so that all and only mechanically viable slip surfaces are considered. In this work, the values of the entry and exit abscissae, $x_{in}$ and $x_{out}$, are bound within the following realistic ranges $x_{in} \in [B, B + \max(H, B)]$ and $x_{out} \in [-\max(H,B), \frac{1}{4}B]$. As for the tangent angle $\delta$, the imposed range is $\delta \in [\delta_{min}, 90°]$ where the upper bound is the right angle, because any larger value would result in a physically unrealistic shear surface, while the lower bound, $\delta_{min}$ depends on the selected pair of entry and exit abscissae $x_{in}$ and $x_{out}$. If $x_{out}$ is located on the scarp, then $\delta_{min} = \tan^{-1}\left(\frac{y_{in}-y_{out}}{x_{in}-x_{out}}\right)$ because, for $\delta < \delta_{min}$, the slip circle becomes unviable due to negative curvature (i.e. for $\delta = \delta_{min}$, the slip circle degenerates into a straight line). Instead, if $x_{out}$ is located on the lower horizontal ground, $\delta_{min} = 2\tan^{-1}\left(\frac{y_{in}-y_{out}}{x_{in}-x_{out}}\right) + \sin^{-1}\left(\frac{x_{in}-x_{out}}{2R}\right)$ because, for $\delta < \delta_{min}$, the slip circle does not entirely lie inside the ground (i.e. for $\delta = \delta_{min}$, the slip circle of radius $R$ passes through the entry point, the exit point and the toe of the slope).

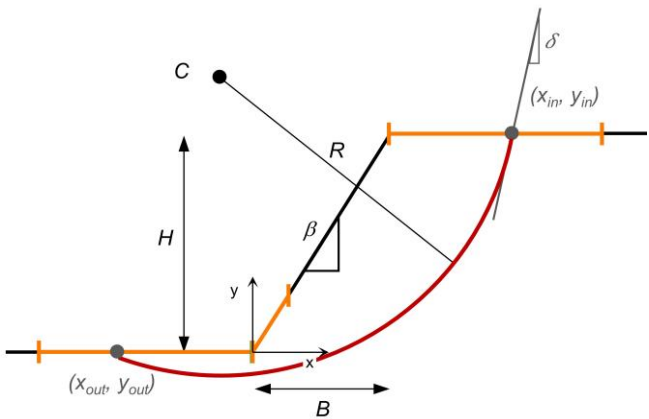

Figure 2. Proposed slip surface parametrisation.

Note that other geometrical variables could be chosen instead of the tangent angle $\delta$ (Figure 3a) or the radius $R$ (Figure 3b) such as, for example, the distance $d$ of the slip surface from the chord connecting $x_{in}$ and $x_{out}$ (Figure 3c) or the distance of the circle centre $(x_c, y_c)$ from the chord connecting $x_{in}$ and $x_{out}$ (Figure 3d). Nevertheless, among all these possibilities, the tangent angle $\delta$ remains the best choice as it directly relates to the physical viability of the slip surface and can therefore be bound by suitable limit values.

Moreover, in discrete (i.e. grid-based) search algorithms, the radius (Figure 3b) and the distance of the centre from the chord (Figure 3d) have the disadvantage of generating an uneven sampling space because equally spaced values of these two parameters lead to a concentration of slip arcs close to the chord.

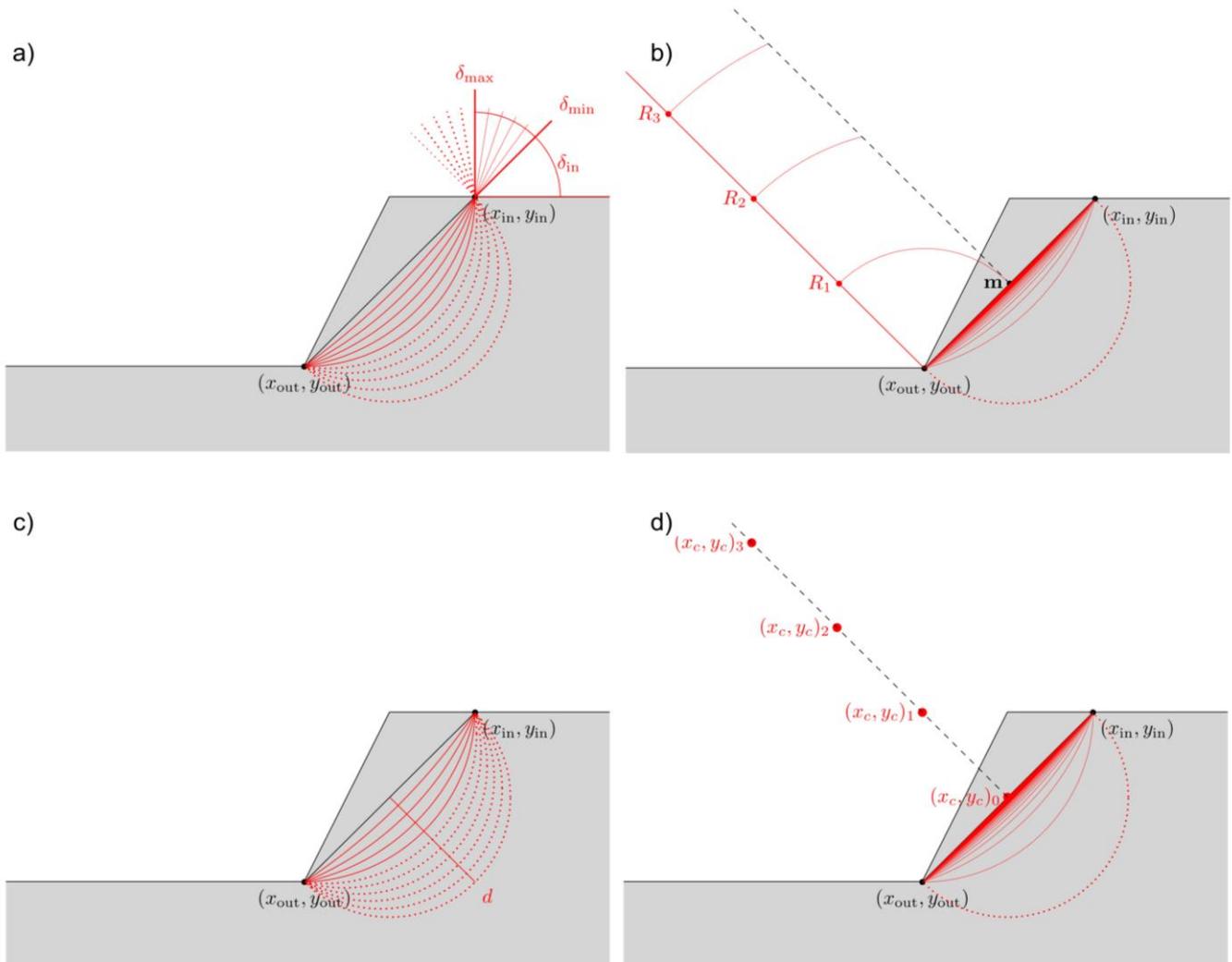

Figure 3. Alternative choices of third parameter: a) tangent angle at the entry point of the slip surface, b) radius of the slip surface, c) distance of the slip surface from the chord and d) distance of the centre from the chord.

## 4 Hybrid grid-simplex algorithm

In discrete (i.e. grid-based) search algorithms, the critical slip surface corresponds to the point in the parametric grid that yields the lowest factor of safety, which implies that the accuracy of the result depends on the coarseness of the discretisation. These potential errors may be overcome by using continuous, rather than discrete, algorithms, which ensure an "unbroken" exploration of the search space. The most popular continuous algorithms require, however, the computation of the gradient of the function to be minimised (Nocedal and Wright, 2006), which is hard to achieve in slope stability analyses. Other gradient-free continuous algorithms, such as the simplex, coordinate descent, trust region, line search, simulated annealing and heuristic algorithms (see Conn et al., 2009; Maurice, 2010; Rios and Sahinidis, 2013; Mirjalili, 2019), have instead the disadvantage of potentially terminating at a local minimum, thus missing the global minimum of the factor of safety.

To circumvent both these limitations, this study employs a hybrid strategy where the entire parametric search space is first explored with a coarse grid to broadly identify the area which is likely to contain the global minimum. Then, a continuous gradient-free search algorithm is employed to locally refine the exploration around the point of the coarse grid with the lowest factor of safety. Among all gradient-free continuous algorithms, the simplex method (also known as Nelder-Mead method) has here been chosen for this local search because it is already coded as an in-built routine inside Python (module *scipy.optimize*), which is the programming environment used in this study.

## 5   Comparison with discrete grid-based search algorithms

This section compares the computational performance of three search algorithms: a) a hybrid algorithm with the improved parametrisation proposed in this work (HI), b) a fine grid algorithm with the improved parametrisation proposed in this work (FI) and c) a fine grid algorithm with standard centre and radius parametrisation (FS). The HI algorithm uses a simplex optimisation to explore the most promising region of the search space, which has been previously identified by a coarse grid with 3 equally spaced values of $x_{in}$, 4 equally spaced values of $x_{out}$ and 5°-spaced values of $\delta$ within their respective bounds. The FI algorithm uses instead a fine grid of 8 equally spaced values of $x_{in}$, 12 equally spaced values of $x_{out}$ and 5°-spaced values of $\delta$ within their respective bounds. Finally, the FS algorithm uses a fine regular grid of 10x10 centres and 10 different equally spaced radii. The comparison is performed for all combinations of the geometric and mechanical parameters listed in Table 1.

Table 1. Geometric and mechanical parameters.

| Slope height $H$ (m) | Slope inclination $\beta$ (°) | Soil unit weight $\gamma$ (kN/m³) | Cohesion $c$ (kPa) | Friction angle $\varphi$ (°) |
|---|---|---|---|---|
| 5 | 10, 20, 30, 40, 50, 60, 70, 80, 90 | 18 | 0.5, 5, 10, 15, 20 | 20, 25, 30, 35, 40 |

### 5.1   Efficiency gains

Figure 4 shows the efficiency gain of the HI and FI algorithms with respect to the FS algorithm for the different values of the slope inclination $\beta$. The efficiency gain is defined as:

$$\text{efficiency gain} = \frac{t_{\text{FS}} - t_{\text{HI/FI}}}{t_{\text{FS}}} \times 100 \qquad (3)$$

where $t_{\text{HI/FI}}$ is the total runtime of either the HI or FI algorithm, i.e. the cumulative runtime for all combinations of cohesion $c$ and friction angle $\varphi$ in Table 1, while $t_{\text{FS}}$ is the reference total runtime of the FS algorithm. The HI algorithm exhibits significant efficiency gains with respect to the FS algorithm, ranging from 80% to 92% as $\beta$ grows from 10° to 90°.

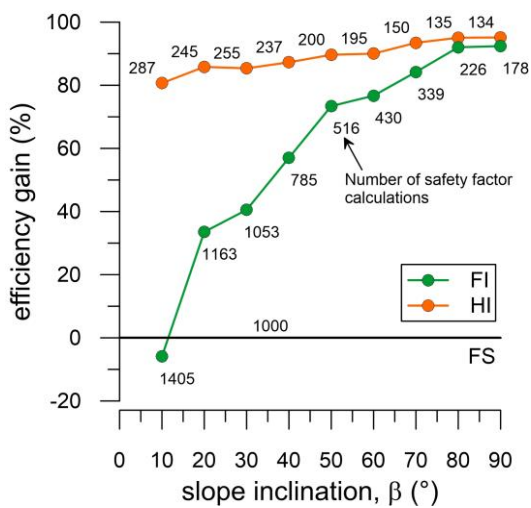

Figure 4. Efficiency gain of FI and HI algorithms with reference to FS algorithm.

The FS algorithm always performs 1000 calculations of the factor of safety (due to the chosen grid of 10x10 centres and 10 radii for every centre) for each triplet of $\beta$, $c$ and $\varphi$. Instead, the number of calculations of the HI algorithm varies not only with the slope inclination $\beta$ but also with the cohesion $c$ and the friction angle $\varphi$. This is because, for each slope inclination $\beta$, the chosen values of cohesion $c$ and friction angle $\varphi$ may lead to a better or worse approximation of the critical slip surface by the initial coarse grid and, hence, to a different number of calculations during the subsequent simplex exploration. The average number of calculations of the HI algorithm for each slope inclination $\beta$ (computed across all pairs of cohesion $c$ and friction angle $\varphi$ in Table 1) is shown in Figure 4 and is considerably smaller than the 1000 calculations of the FS algorithm, which largely explains the observed efficiency gains.

Figure 4 also shows the relatively good performance of the FI algorithm with respect to the FS algorithm. Also in this case, the efficiency gain is mostly caused by the lower number of calculations and grows as the slope inclination $\beta$ increases up to 92% for $\beta = 90°$. Unlike the FS algorithm, the FI algorithm performs a variable number of calculations as the bounds of parameter $\delta$ (and hence the corresponding number of sampling points of the grid) depends on the slope inclination $\beta$. Figure 4 shows that the number of calculations of the FI algorithm reduces as the slope inclination $\beta$ increases, down to a minimum of 178 for $\beta = 90°$. Only for $\beta = 10°$, the FI algorithm is slightly slower than the FS algorithm because the number of calculations of the FI algorithm, i.e. 1405, is higher than that of the FS algorithm, i.e. 1000. Interestingly, for $\beta = 20°$ and $\beta = 30°$, the number of calculations of the FI algorithm is still higher than 1000 (i.e. 1163 and 1053, respectively), yet the runtime is smaller than that of the FS algorithm. This apparently counterintuitive result is explained by the recursive calculation of the factor of safety $F$ in Equation (2), which requires less iterations to converge for physically consistent slip surfaces.

As an aside, Figure 5 shows, for each slope inclination $\beta$, the number of calculations by the simplex optimisation in the HI algorithm when the cohesion $c$ and friction angle $\varphi$ change. Note that, to produce a smooth plot, Figure 5 uses a finer discretisation of cohesion and friction angle than that of Table 1. Inspection of Figure 5 indicates that the number of calculations of the safety factor is largest when $\beta = 60°$, though there is no clear correlation with the values of cohesion and friction angle.

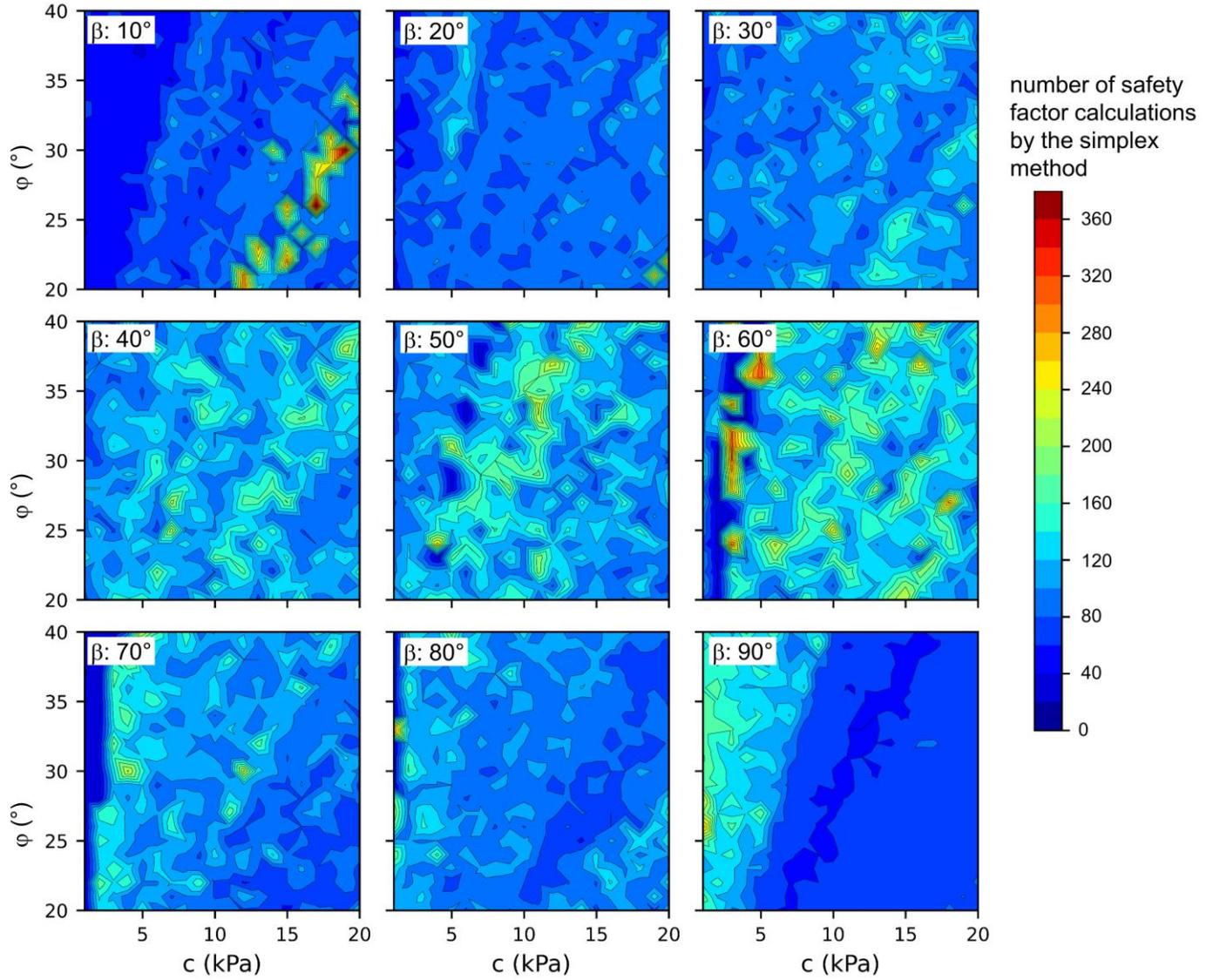

Figure 5. Number of calculations of the safety factor by the simplex method as cohesion $c$ and friction angle $\varphi$ change for different slope inclinations $\beta$.

## 5.2 Accuracy gains

The relative accuracy of the factor of safety computed by the three algorithms, i.e. $F_{FS}$, $F_{FI}$ and $F_{HI}$, for each combination of $\beta$, $c$ and $\varphi$ is compared by plotting the frequency histograms of the ratios $F_{FI}/F_{FS}$ and $F_{HI}/F_{FS}$ (Figure 6). Both FI and HI algorithms are more accurate than the FS algorithm as they consistently calculate smaller values of the factor of safety, with a maximum increase in accuracy of about 25%. The medians of the two histograms of $F_{FI}/F_{FS}$ and $F_{HI}/F_{FS}$ are respectively 96.1% and 95.7%, i.e. 50% of the runs of the FI and HI algorithms are at least 3.9% and 4.3% more accurate than the FS algorithm, respectively. Instead, the mean and mode are 95.3% and 96.1% for the $F_{FI}/F_{FS}$ histogram while they are 94.9% and 95.7% for the $F_{HI}/F_{FS}$ histogram. In less than 10 instances out of a total of 225 analyses, the values of $F_{FI}$ or $F_{HI}$ are larger than the value of $F_{FS}$ but, even then, the ratio is at most 100.8%, which is practically insignificant. Figure 6 also shows that the $F_{HI}/F_{FS}$ histogram is marginally flatter and slightly shifted to the left compared to the $F_{FI}/F_{FS}$ histogram, which means that the HI algorithm is slightly more accurate than the FI algorithm. Both HI and FI algorithms exploit the novel geometrical parametrisation, so the slightly greater accuracy of the HI algorithm compared to the FI algorithm must be attributed to the continuous simplex optimisation in the former case.

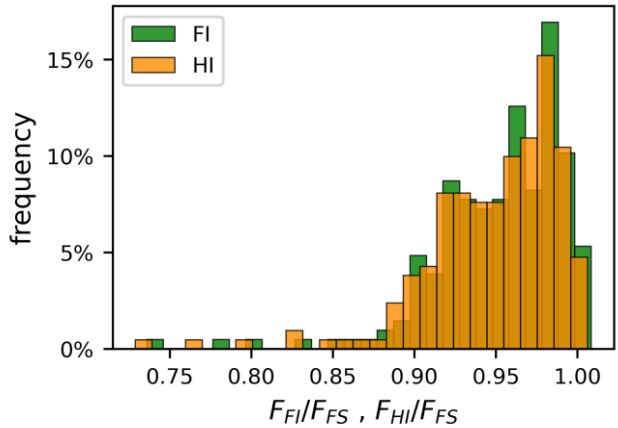

Figure 6. Accuracy gains of FI and HI algorithms with reference to FS algorithm.

## 6  Comparison with heuristic algorithms

Several authors have already used continuous gradient-free optimisation algorithms for the minimisation of Equation (2) to calculate the factor of safety according to Bishop (1955). Among these algorithms, particle swarm optimisation (PSO) and genetic algorithms (GA) have proved most accurate and least prone to attain local minima, though they are undermined by high computational costs. To provide a comprehensive and fair comparison, Table 2 evaluates both the accuracy and computational efficiency of the HI algorithm of this work against the PSO algorithm by Kalatehjari *et al.* (2012) and Himansu and Burman (2019), as well as the GA algorithm by Zolfaghari *et al.* (2005). Note that all these authors adopt a conventional parametrisation of the slip circles, which is based on the definition of the corresponding centres and radii. The comparison is made for the four different benchmark cases reported in Himansu and Burman (2019), i.e. two homogeneous slopes and two layered ones, all having the same inclination $\beta = 26.56°$ but different heights and/or material properties as listed in Table 2.

Inspection of Table 2 confirms the higher efficiency of the HI algorithm, which requires approximately 300 calculations of the factor of safety to solve each case. Conversely, the other algorithms require from 3000 to 10000 calculations, respectively, i.e. about 10 to 30 times more. As anticipated, the higher computational efficiency of the HI algorithm is due to: i) the improved parametrisation of the slipe circles, which restricts the search space to all and only viable surfaces, and ii) the hybrid search algorithm, which allows for the continuous exploration of only the most promising region of the search space. Note that the preliminary coarse grid search always performs 146 calculations for all four slopes because of their identical inclination. Therefore, the varying numbers of calculations between the four cases in Table 2 are solely attributed to the differences during the local simplex exploration.

In terms of accuracy, the values of the factor of safety calculated by the HI algorithm are in excellent agreement with those calculated by the PSO and GA. In particular, the HI algorithm is slightly more conservative than the GA by Zolfaghari *et al.* (2005) as it predicts a marginally smaller factor of safety. Conversely, the HI algorithm is slightly less conservative than the PSO of Kalatehjari *et al.* (2012) and Himanshu and Burman (2019) as it predicts a marginally larger factor of safety. Nevertheless, the differences between all factors of safety do not exceed 2%, which is irrelevant if weighted against the massively different number of calculations.

The HI algorithm therefore represents an efficient compromise between accuracy and computational effort, which is of crucial importance when dealing with large numbers of calculations, such as during the statistical treatment of material and physical uncertainties at the regional scale. Table 2 also shows that the discretisation of the unstable soil mass in 25 slices is sufficiently fine as doubling the number of slices results in a negligible change of the value of the factor of safety.

Finally, Figure 7 compares the critical slip lines calculated by the HI algorithm and the PSO of Himansu and Burman (2019), which further confirms the excellent agreement between the two solutions. For the sake of clarity, the critical slip lines calculated by the PSO of Kalatehjari *et al.* (2012) and the GA of Zolfaghari *et al.* (2005) have been omitted.

Table 2. Comparison of computational efficiency and accuracy of HI algorithm versus PSO and GA.

| Case | Geometry | Soil parameters | Algorithm type | N. of slices | N. of calculations | Factor of safety |
|---|---|---|---|---|---|---|
| 1 | $H$=5m $B$=10m $\beta$=26.56° | $c$=9.8kPa $\varphi$=10° $\gamma$=17.64 kN/m³ | PSO, Kalatehjari *et al.* (2012) | 24 | 3500 | 1.3128 |
| | | | PSO, Himanshu and Burman (2019) | 27 | 10000 | 1.3136 |
| | | | **HI algorithm, this study** | **25** | **294** | **1.3429** |
| | | | **HI algorithm, this study** | **50** | **255** | **1.3426** |
| 2 | $H$=8.5m $B$=17m $\beta$=26.56° | $c$=14.71kPa $\varphi$=20° $\gamma$=18.63 kN/m³ | GA, Zolfaghari *et al.* (2005) | 150 | 3000 | 1.74 |
| | | | PSO Kalatehjari *et al.* (2012) | 40 | 3500 | 1.7197 |
| | | | PSO, Himanshu and Burman (2019) | 42 | 10000 | 1.7195 |
| | | | **HI algorithm, this study** | **25** | **286** | **1.7336** |
| | | | **HI algorithm, this study** | **50** | **258** | **1.7363** |
| 3 | $H$=5m $B$=10m $\beta$=26.56° | layer 1: $c$=14.71kPa $\varphi$=20° $\gamma$=18.63 kN/m³ layer 2: $c$=9.8kPa $\varphi$=10° $\gamma$=17.64 kN/m³ | PSO, Himanshu and Burman (2019) | 42 | 10000 | 1.3395 |
| | | | **HI algorithm, this study** | **25** | **322** | **1.3645** |
| 4 | $H$=5m $B$=10m $\beta$=26.56° | layer 1: $c$=14.71kPa $\varphi$=20° $\gamma$=18.63 kN/m³ layer 2: $c$=9.8kPa $\varphi$=10° $\gamma$=17.64 kN/m³ | PSO, Himanshu and Burman (2019) | 42 | 10000 | 1.3183 |
| | | | **HI algorithm, this study** | **25** | **272** | **1.3438** |

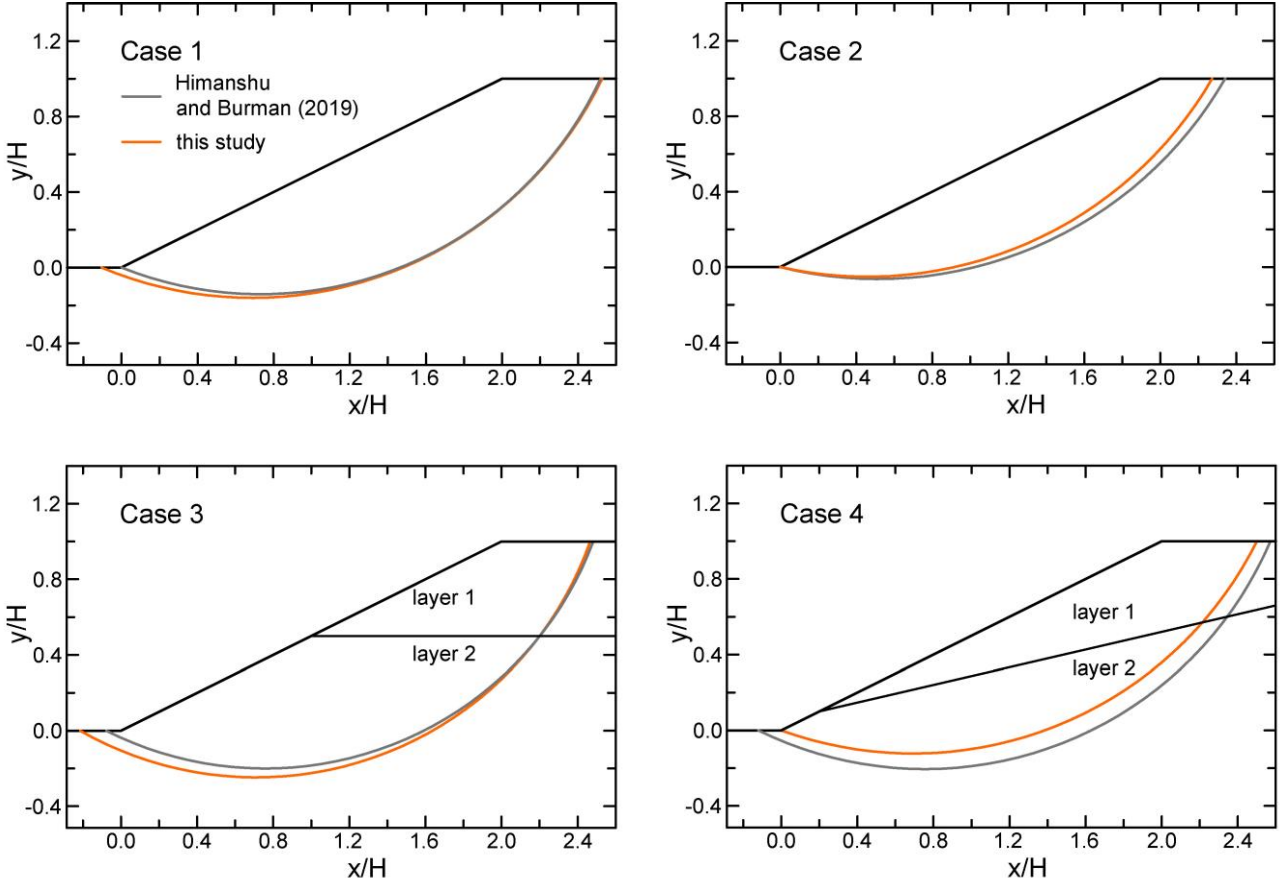

Figure 7. Critical slip surfaces of the four benchmark cases in Table 2.

## 7   Comparison with upper bound solutions

The method of slices is widely used in engineering practice because of its simplicity, accuracy and compliance with existing design norms but it is theoretically less rigorous than other approaches such as upper bounds limit analyses (Chen, 1975). These latter approaches have demonstrated that the assumption of a log-spiral failure mechanism, rather than a circular one, is theoretically more robust for the study of cohesive-frictional homogeneous slopes. In this respect, Huang (2023) recently presented an upper bound stability analysis of homogenous slopes accounting for the variation of both geometrical ($H$, $\beta$) and mechanical ($\gamma$, $c$, $\varphi$) parameters while assuming a log-spiral failure mechanism.

Therefore, to explore the relative validity of the Bishop (1955) HI algorithm proposed in this work, Table 3 presents a comparison with the upper bound analyses performed by Huang (2023) over a range of geometrical and mechanical parameters. For each slope inclination $\beta$, a total of 30 different cases have been compared corresponding to multiple combinations of geometrical ($H$, $\beta$) and mechanical ($\gamma$, $c$, $\varphi$) parameters. Table 3 reports, for each slope inclination $\beta$, the mean, median and maximum absolute percentage difference of the factor of safety computed by the HI algorithm with respect to the upper bound analysis of Huang (2023). Inspection of Table 3 indicates that the maximum values of the mean, median and maximum absolute percentage differences across all slope inclinations are 1.38%, 1.76% and 5.19%, respectively. This demonstrates a very good agreement of the HI algorithm, implementing Bishop (1955) method of slices, with the theoretically more rigorous upper bound limit analysis, which further increases the attractiveness of the proposed approach.

Table 3. Comparison of Bishop (1955) HI algorithm with Huang (2023) upper bound limit analysis.

| beta (°) | absolute mean difference (%) | absolute median difference (%) | absolute maximum difference (%) |
|---|---|---|---|
| 15 | 1.13 | 1.74 | 4.59 |
| 20 | 0.58 | 1.19 | 5.14 |
| 25 | 0.05 | 0.60 | 5.19 |
| 30 | 0.40 | 0.03 | 4.70 |
| 35 | 0.77 | 0.29 | 3.62 |
| 40 | 1.02 | 0.83 | 2.76 |
| 45 | 1.14 | 1.30 | 2.48 |
| 50 | 1.11 | 1.76 | 3.59 |
| 55 | 0.33 | 0.30 | 1.79 |
| 60 | 0.22 | 0.25 | 1.62 |
| 65 | 0.68 | 0.70 | 2.25 |
| 70 | 1.04 | 1.10 | 2.76 |
| 75 | 1.28 | 1.39 | 3.14 |
| 80 | 1.38 | 1.54 | 3.35 |
| 85 | 1.31 | 1.54 | 3.38 |
| 90 | 1.05 | 1.39 | 3.20 |
| **Overall** | **1.38*** | **1.76*** | **5.19*** |

* maximum value

## 8 Conclusions and future work

This paper has presented an optimised search algorithm to identify the critical slip surface corresponding to the lowest value of the factor of safety in slope stability analyses. The algorithm introduces a novel parametrisation of slip surfaces, which are defined by the entry point into the slope, the exit point out of the slope and the tangent angle at the entry point. These parameters are clearly linked to the physical characteristics of the corresponding slip surfaces and are therefore easily bound to restrict the search to all and only viable mechanisms. The search algorithm first computes the factor of safety over a coarse grid covering the entire parameter space to identify the region characterised by lower values of the factor of safety, which is then finely explored via a local continuous search by means of a simplex optimisation.

This hybrid search strategy and the physically based parametrisation of slip surfaces reduce computation time by about 90% compared to conventional approaches. The proposed approach also increases accuracy by about 5% on average, with a maximum of about 25%, as the proposed algorithm can pick slip surfaces that are not detected by conventional grid-based searches, which are instead restricted to a set of pre-generated failure geometries. The proposed hybrid search algorithm with physically based slip parametrisation is also more efficient than recently proposed heuristic algorithms, such as particle swarm optimisation and genetic algorithms, while preserving similar levels of accuracy. Finally, the proposed method shows excellent agreement with the upper bound solution of Huang (2023), which is based on the log-spiral failure mechanism.

In forthcoming work, the above gains of efficiency will be exploited within computationally demanding GIS models to produce landslide susceptibility maps at basin scale, where the statistical treatment of uncertainties requires the evaluation of the factor of safety for numerous combinations of mechanical and geometrical parameters. The proposed algorithm will also be extended to consider non-circular slip surfaces and to incorporate different versions of the method of slices as proposed in the literature.

## Acknowledgements
This study was funded by the European Union - NextGenerationEU and by the Ministry of University and Research (MUR), National Recovery and Resilience Plan (NRRP), Mission 4, Component 2, Investment 1.5, project "RAISE - Robotics and AI for Socio-economic Empowerment" (ECS00000035). Andrea Bressan, Simone Pittaluga, Lorenzo Tamellini and Domenico Gallipoli are part of RAISE Innovation Ecosystem.

## CRediT authorship contribution statement
**Leonardo Maria Lalicata**: Conceptualisation, Methodology, Validation, Formal analysis, Investigation, Data curation, Writing – original draft, Visualisation. **Andrea Bressan**: Conceptualisation, Methodology, Validation, Formal analysis, Investigation, Writing – Review & Editing, Visualisation. **Simone Pittaluga**: Writing – Review & Editing. **Lorenzo Tamellini**: Methodology, Formal analysis, Writing – Review & Editing, Supervision. **Domenico Gallipoli**: Methodology, Formal analysis, Writing – Review & Editing, Supervision, Project administration, Funding acquisition.

## References

Baker, R. (1980). Determination of the critical slip surface in slope stability computations. *International Journal for Numerical and Analytical Methods in Geomechanics*, *4*(4), 333-359.

Bell, J. M. (1968). General Slope Stability Analysis. *Journal of the Soil Mechanics and Foundations Division*, *94*(6), 1253–1270. https://doi.org/10.1061/JSFEAQ.0001196

Bishop, A. W. (1955). The use of the Slip Circle in the Stability Analysis of Slopes. *Géotechnique*, *5*(1), 7–17. https://doi.org/10.1680/geot.1955.5.1.7

Bishop, A. W., & Morgenstern, N. (1960). Stability Coefficients for Earth Slopes. *Géotechnique*, *10*(4), 129–153. https://doi.org/10.1680/geot.1960.10.4.129

Bond, A., J., Schuppener, B., Scarpelli, G., & Orr, T.L.L. (2013). Eurocode 7 geotechnical design: worked examples. *Joint Research Centre, Institute for the Protection and Security of the Citizen.* Dimova, S. (editor), Pinto, A. (editor), Nikolova, B. (editor), Publications Office, 2013, https://data.europa.eu/doi/10.2788/3398

Boutrup, E., & Lovell, C. W. (1980). Searching techniques in slope stability analysis. *Engineering Geology*, *16*(1), 51–61. https://doi.org/10.1016/0013-7952(80)90006-X

Cascini, L., Cuomo, S., & Sorbino, G. (2005). Flow-like mass movements in pyroclastic soils: Remarks on the modelling of triggering mechanisms. *Italian Geotechnical Journal*, *4*, 11–31.

Celestino, T. B., and Duncan, J. M. (1981). Simplified search for noncircular slip surface. *Proceeding of 10$^{th}$ International Conference on Soil Mechanics and Foundation Engineering*, A. A. Balkema, Rotterdam, The Netherlands, 3, 391-394.

Chen, W. F. 1975. *Limit analysis and soil plasticity*. Amsterdam, Netherlands: Elsevier Science.

Cheng, Y. M., Li, L., Chi, S., & Wei, W. B. (2007). Particle swarm optimization algorithm for the location of the critical non-circular failure surface in two-dimensional slope stability analysis. *Computers and Geotechnics, 34*(2), 92–103. https://doi.org/10.1016/j.compgeo.2006.10.012

Conn, A. R., Scheinberg, K., & Vicente, L. N. (2009). *Introduction to derivative-free optimization*. Society for Industrial and Applied Mathematics.

Cotecchia, F., Tagarelli, V., Pedone, G., Ruggieri, G., Guglielmi, S., & Santaloia, F. (2019). Analysis of climate-driven processes in clayey slopes for early warning system design. *Proceedings of the Institution of Civil Engineers-Geotechnical Engineering*, 172(6), 465-480.

Duncan, J. M. (1996). State of the Art: Limit Equilibrium and Finite-Element Analysis of Slopes. *Journal of Geotechnical Engineering*, *122*(7), 577–596. https://doi.org/10.1061/(ASCE)0733-9410(1996)122:7(577)

EN 1997-1 (2004). Eurocode 7: Geotechnical design - Part 1: General rules. *The European Union per Regulation 305/2011, Directive 98/34/EC, Directive 2004/18/EC*.

Federici, B., Bovolenta, R., & Passalacqua, R. (2015). From rainfall to slope instability: An automatic GIS procedure for susceptibility analyses over wide areas. *Geomatics, Natural Hazards and Risk*, *6*(5–7), 454–472. https://doi.org/10.1080/19475705.2013.877087

Firincioglu, B. S., & Ercanoglu, M. (2021). Insights and perspectives into the limit equilibrium method from 2D and 3D analyses. *Engineering Geology*, 281, 105968. https://doi.org/10.1016/j.enggeo.2020.105968

Govi, M., Mortara, G., & Sorzana, P. F. (1985). Eventi idrologici e frane. *Eventi idrologici e frane*, *20*, 359–375.

Himanshu, N., & Burman, A. (2019). Determination of Critical Failure Surface of Slopes Using Particle Swarm Optimization Technique Considering Seepage and Seismic Loading. *Geotechnical and Geological Engineering*, *37*(3), 1261–1281. https://doi.org/10.1007/s10706-018-0683-8

Himanshu, N., Kumar, V., Burman, A., Maity, D., & Gordan, B. (2021). Grey wolf optimization approach for searching critical failure surface in soil slopes. *Engineering with Computers*, 37(3), 2059–2072. https://doi.org/10.1007/s00366-019-00927-6

Huang, W. (2023). Stability of Homogeneous Slopes: From Chart to Closed-Form Solutions and from Deterministic to Probabilistic Analysis. International Journal of Geomechanics, 23(9), 04023136.

ISPRA- Istituto Superiore per la Protezione e la Ricerca Ambientale. 2023. Nota: quadro di sintesi dissesto frane Emilia-Romagna, Italia ((aggiornamento 22/05/2023). "https://www.isprambiente.gov.it/news"

Jaedicke, C., Van Den Eeckhaut, M., Nadim, F., Hervás, J., Kalsnes, B., Vangelsten, B. V., Smith, J. T., Tofani, V., Ciurean, R., Winter, M. G., Sverdrup-Thygeson, K., Syre, E., & Smebye, H. (2013). Identification of landslide hazard and risk 'hotspots' in Europe. *Bulletin of Engineering Geology and the Environment*. https://doi.org/10.1007/s10064-013-0541-0

Kalatehjari, R., Ali, N., Hajihassani, M., & Fard, M. K. (2012). The Application of Particle Swarm Optimization in Slope Stability Analysis of Homogeneous Soil Slopes. *International Review on Modelling and Simulations*, 5(1), 458-465.

Le, T. M. H., Gallipoli, D., Sánchez, M., & Wheeler, S. (2015). Stability and failure mass of unsaturated heterogeneous slopes. *Canadian Geotechnical Journal*, 52(11), 1747-1761.

Mirjalili, S. (2019). *Genetic algorithm*. Evolutionary Algorithms and Neural Networks: Theory and Applications, 43-55.

Montrasio, L., & Valentino, R. (2008). A model for triggering mechanisms of shallow landslides. *Natural Hazards and Earth System Sciences*, *8*(5), 1149–1159. https://doi.org/10.5194/nhess-8-1149-2008

Morgenstern, N. R., & Price, V. E. (1965). The Analysis of the Stability of General Slip Surfaces. *Géotechnique*, *15*(1), 79–93. https://doi.org/10.1680/geot.1965.15.1.79

Nguyen, V. U. (1985). Determination of critical slope failure surfaces. *Journal of Geotechnical Engineering*, *111*(2), 238-250.

Nocedal, J., Wright, S.J. (2006). *Numerical optimization*. Spinger.

Notti, D., Wrzesniak, A., Dematteis, N., Lollino, P., Fazio, N. L., Zucca, F., & Giordan, D. (2021). A multidisciplinary investigation of deep-seated landslide reactivation triggered by an extreme rainfall event: a case study of the Monesi di Mendatica landslide, Ligurian Alps. *Landslides*, *18*, 2341-2365.

Pedone, G., Ruggieri, G., & Trizzino, R. (2018). Characterisation of climatic variables used to identify instability thresholds in clay slopes. *Géotechnique Letters*, 8(3), 231-239.

Rahardjo, H., Zhai, Q., Satyanaga, A. *et al.* (2023). Slope susceptibility map for preventive measures against rainfall-induced slope failure. *Urban Lifeline* 1(5). https://doi.org/10.1007/s44285-023-00006-9

Rios, L. M., & Sahinidis, N. V. (2013). Derivative-free optimization: a review of algorithms and comparison of software implementations. *Journal of Global Optimization*, 56, 1247-1293.

Romeo, S., D'Angiò, D., Fraccica, A., Licata, V., Vitale, V., Chiessi, V., Amanti, M., & Bonasera, M. (2023). Investigation and preliminary assessment of the Casamicciola landslide in the island of Ischia (Italy) on November 26, 2022. *Landslides*, *20*(6), 1265–1276. https://doi.org/10.1007/s10346-023-02064-0

Sarma, S. K. (1973). Stability analysis of embankments and slopes. *Géotechnique*, *23*(3), 423–433. https://doi.org/10.1680/geot.1973.23.3.423

Sarma, S. K. (1979). Stability Analysis of Embankments and Slopes. *Journal of the Geotechnical Engineering Division*, *105*(12), 1511–1524. https://doi.org/10.1061/AJGEB6.0000903

Siegel, R. A., Kovacs, W. D., & Lovell, C. W. (1981). Random surface generation in stability analysis. *Journal of the Geotechnical Engineering Division*, *107*(7), 996-1002.

SLIDE2– *[Computer software]*. Roc-science International Ltd, Toronto, ON, Canada

SLOPE/W (2007). *GeoStudio [Computer software]*. GEO/SLOPE International Ltd, Calgary, AB, Canada.

Spencer, E. (1967). A Method of analysis of the Stability of Embankments Assuming Parallel Inter-Slice Forces. *Géotechnique*, *17*(1), 11–26. https://doi.org/10.1680/geot.1967.17.1.11

Spencer, E. (1973). Thrust line criterion in embankment stability analysis. *Géotechnique*, *23*(1), 85–100. https://doi.org/10.1680/geot.1973.23.1.85

Steward, T., Sivakugan, N., Shukla, S. K., & Das, B. M. (2011). Taylor's Slope Stability Charts Revisited. *International Journal of Geomechanics*, *11*(4), 348–352. https://doi.org/10.1061/(ASCE)GM.1943-5622.0000093

Trigila, A., Iadanza, C., & Spizzichino, D. (2010). Quality assessment of the Italian Landslide Inventory using GIS processing. *Landslides*, *7*(4), 455–470. https://doi.org/10.1007/s10346-010-0213-0

Versace, P., Sirangelo, B., and Capparelli, G. (2002). A hydrological model to assess the activation probability of landslides triggered by rainfall, *Proceedings of 1st Italian-Russian Workshop "New Trends in Hydrology"*, Rende (CS), Italy.

Zhu, D. Y. (2008). Investigations on the accuracy of the simplified Bishop method. In *Landslides and Engineered Slopes. From the Past to the Future*, pp. 1077-1080. Chen, Zhang, Ho, Wu, Li Eds. CRC Press.

Zolfaghari, A. R., Heath, A. C., & McCombie, P. F. (2005). Simple genetic algorithm search for critical non-circular failure surface in slope stability analysis. *Computers and Geotechnics*, *32*(3), 139–152.